\documentclass[aps,pra,twocolumn,showpacs,amsmath,amssymb,preprintnumbers,superscriptaddress,10pt,longbibliography]{revtex4-1}

\usepackage{bm}
\usepackage{graphicx}
\graphicspath{{figures/}}
\usepackage{SIunits}
\usepackage{hyphenat}
\usepackage{physics}
\usepackage{scrextend}
\usepackage[bookmarks=false]{hyperref}
\usepackage{color}
\usepackage[capitalise]{cleveref}
\crefname{section}{Sec.}{Secs.}
\Crefname{section}{Section}{Sections}

\definecolor{pink}{RGB}{255,0,255}

\definecolor{red}{rgb}{0,0,1}

\definecolor{green}{RGB}{0,200,0}

\begin{document}

\title{Laser seeding attack in quantum key distribution}

\author{Anqi~Huang}
\email{angelhuang.hn@gmail.com}
\affiliation{Institute for Quantum Information \& State Key Laboratory of High Performance Computing, College of Computer, National University of Defense Technology, Changsha 410073, People's Republic of China}
\affiliation{Institute for Quantum Computing, University of Waterloo, Waterloo, ON, N2L~3G1 Canada}

\author{{\'A}lvaro~Navarrete}
\affiliation{EI Telecomunicaci\'on, Department of Signal Theory and Communications, University of Vigo, Vigo E-36310, Spain} 

\author{Shi-Hai~Sun}
\affiliation{School of Physics and Astronomy, Sun Yat-Sen University, Zhuhai 519082, People's Republic of China}

\author{Poompong~Chaiwongkhot}
\affiliation{Institute for Quantum Computing, University of Waterloo, Waterloo, ON, N2L~3G1 Canada}
\affiliation{Department of Physics and Astronomy, University of Waterloo, Waterloo, ON, N2L~3G1 Canada}

\author{Marcos~Curty}
\affiliation{EI Telecomunicaci\'on, Department of Signal Theory and Communications, University of Vigo, Vigo E-36310, Spain} 

\author{Vadim~Makarov}
\affiliation{Russian Quantum Center, Skolkovo, Moscow 121205, Russia}
\affiliation{\mbox{Shanghai Branch, National Laboratory for Physical Sciences at Microscale and CAS Center for Excellence in} \mbox{Quantum Information, University of Science and Technology of China, Shanghai 201315, People's Republic of China}}
\affiliation{NTI Center for Quantum Communications, National University of Science and Technology MISiS, Moscow 119049, Russia}
\affiliation{Department of Physics and Astronomy, University of Waterloo, Waterloo, ON, N2L~3G1 Canada}

\begin{abstract}
Quantum key distribution (QKD) based on the laws of quantum physics allows the secure distribution of secret keys over an insecure channel. Unfortunately, imperfect implementations of QKD compromise its information-theoretical security. Measurement-device-independent quantum key distribution (MDI-QKD) is a promising approach to remove all side channels from the measurement unit, which is regarded as the ``Achilles' heel'' of QKD. An essential assumption in MDI-QKD is however that the sources are trusted. Here we experimentally demonstrate that a practical source based on a semiconductor laser diode is vulnerable to a laser seeding attack, in which light injected from the communication line into the laser results in an increase of the intensities of the prepared states. The unnoticed increase of intensity may compromise the security of QKD, as we show theoretically for the prepare-and-measure decoy-state BB84 and MDI-QKD protocols. Our theoretical security analysis is general and can be applied to any vulnerability that increases the intensity of the emitted pulses. Moreover, a laser seeding attack might be launched as well against decoy-state based quantum cryptographic protocols beyond QKD.
\end{abstract}

\maketitle

\section{Introduction}
\label{sec:intro}

The distribution of a secret key between two authorized parties, Alice and Bob, is a fundamental but challenging cryptographic task. Such secret key is the essential resource of the one-time-pad algorithm~\cite{vernam1926}, the only known encryption method that can offer unconditionally secure communications. Public-key cryptography solves this problem by resorting to computational assumptions, for instance, the difficulty of factoring large numbers~\cite{rivest1978}. This approach is however vulnerable to technological advances in both hardware and software; indeed, it is well-known that factoring is an easy problem on a quantum computer~\cite{shor1997}. Quantum key distribution (QKD), on the other hand, provides a solution based on the laws of quantum physics, and thus, in theory, it can guarantee that the distributed keys are information-theoretically secure~\cite{gisin2002,scarani2009,lo2014}. 

There is however a big gap between the theory and the practice of QKD because the behaviour of real QKD devices typically deviates from that considered in the security proofs. Such deviation could be exploited by an eavesdropper, Eve, to obtain information about the secret key without being detected in QKD implementations~\cite{makarov2006,qi2007,lamas-linares2007,lydersen2010a,lydersen2010b,wiechers2011,lydersen2011c,lydersen2011b,gerhardt2011,sun2011,jain2011, bugge2014,sajeed2015a,huang2016,sajeed2016,makarov2016,huang2018,huang2019b,zheng2019,zheng2019a,chistiakov2019}. Most of the quantum hacking attacks realized so far exploit imperfections of the single-photon detectors (SPDs) -- the ``Achilles' heel'' of QKD~\cite{makarov2006,qi2007,lamas-linares2007,lydersen2010a,lydersen2010b,wiechers2011,lydersen2011c,lydersen2011b,gerhardt2011,bugge2014,sajeed2015a,huang2016,makarov2016}. Indeed, in recent years there has been an enormous effort to try to close the detectors' security loopholes and restore the security of QKD realizations. Some solutions are based on hardware and software patches~\cite{lim2015,dixon2017}, whose drawback is however that each patch typically protects only against a specific loophole, {\it i.e.,}\ the system might still be vulnerable to unknown attacks. Moreover, patches might also be hacked~\cite{huang2016,makarov2016}. A safer and more elegant solution is that of measurement-device-independent QKD (MDI-QKD)~\cite{lo2012,curty2014}. Remarkably, this latter approach guarantees security independently of the behaviour of the measurement device, which can be treated as a ``black box'' fully controlled by Eve. This is achieved by turning Bob's receiver into a transmitter by means of a time-reversed Einstein-Podolsky-Rosen~(EPR) protocol~\cite{biham1996,inamori2002}. MDI-QKD has been successfully demonstrated in several recent experiments~\cite{rubenok2013,silva2013,liu2013,tang2014,comandar2016,tang2016} including an implementation over $404$ km~\cite{yin2016}.

With the advent of MDI-QKD all security loopholes from the measurement unit are closed, so the focus is now on how to protect the QKD transmitters. For instance, decoy-state QKD~\cite{hwang2003,wang2005a,lo2005} is a practical solution to defeat the photon-number-splitting attack~\cite{huttner1995,brassard2000}. More recently, several works have considered other imperfections of the transmitter, and new security proofs that guarantee security in the presence of such imperfections have been developed~\cite{tamaki2014,mizutani2015,lucamarini2015,tamaki2016,wang2018,yoshino2018,mizutani2018,pereira2019}. For example, Refs.~\cite{lucamarini2015,tamaki2016,wang2018} quantify the optical isolation that is needed in order to achieve a certain performance ({\it i.e.,}\ a certain secret key rate over a given distance) in the presence of a Trojan-horse attack~(THA), in which Eve injects bright light into the transmitter and then analyses the back-reflected light to obtain information about the quantum signals emitted. Finally, a type of light injection attack that affects the operation of the laser diode in the transmitter has recently been introduced, allowing Eve to actively de-randomise the source's phase and even change other parameters~\cite{sun2015}. Indeed, the use of non-phase-randomised signals has a severe effect on the security of QKD, as has been shown in the past decade~\cite{lo2007,sun2012,tang2013,sun2014}.

While the results above are promising, there is still a long way to go to be able to ensure the security of QKD implementations. For instance, a fundamental assumption of QKD is that the intensity of the quantum states prepared by Alice is set at a single-photon level. This assumption is indeed vital for a QKD system. However, no study has investigated whether or not Eve could increase the mean photon number of the prepared states. Here we introduce, and experimentally demonstrate, a quantum hacking attack, which we call ``laser seeding attack'', that can {\em increase and control the intensity} of the light emitted by the laser diode in the transmitter of a QKD system. This attack has been confirmed experimentally for two types of laser diodes. Different from the THA that analyses the back-reflected light that is originally from an external independent source, the laser seeding attack manipulates the functioning of the transmitter's laser diode directly. That is, while in a THA Eve tries to correlate her signals with the quantum states prepared by the legitimate users of the system, in a laser seeding attack the goal of Eve is to directly increase the intensity of such quantum states. Most importantly, this attack seriously compromises the security of decoy-state based QKD, which includes MDI-QKD with practical light sources as a prominent example. More precisely, in the presence of this attack, current security analyses overestimate the resulting secret key rate and thus they do not guarantee security. 

\section{Experimental setup}
\label{sec:scheme}

\begin{figure}
	\includegraphics{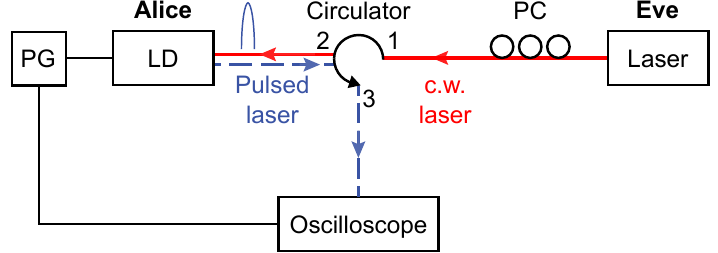}
	\caption{Experimental setup. The red solid arrows represent Eve's injected continuous-wave (c.w.)\ bright light, and the blue dashed arrows indicate the optical pulses emitted by Alice' laser diode. PG: electronic pulse generator; LD: laser diode; PC: polarization controller.}
	\label{fig:scheme}
\end{figure}

To investigate to which extent Eve can increase the output optical power of a laser diode by injecting light into it, we conduct an experiment whose schematic is illustrated in~\cref{fig:scheme}. On Alice's side, the laser diode, as a testing target, generates optical pulses. As a hacker, Eve employs a tunable laser (Agilent 8164B) to send continuous-wave (c.w.)\ bright light to Alice's laser diode via a single-mode optical fibre. The tunable laser is able to adjust the wavelength and output power of the signals emitted so that Eve can inject photons with a proper wavelength into Alice's laser. In so doing, the energy of each injected photon can match the energy difference between the excited state and the ground state of the laser, and thus satisfy the condition for stimulated emission.

In order to maximize the injection efficiency, a polarization controller is used to adjust the polarization of Eve's laser such that it matches that of Alice's laser. To separate Eve's injected light from that emitted by Alice, we employ an optical circulator. Eve's light enters port 1 of the circulator and exits through its port 2, while Alice's light goes from port 2 of the circulator to its port 3 (see~\cref{fig:scheme}). We record Alice's output pulses with an optical-to-electrical converter with $40~\giga\hertz$ bandwidth (Picometrix PT-40A) that is connected to a high-speed oscilloscope (Agilent DSOX93304Q) of $33~\giga\hertz$ bandwidth. The average pulse energy is then calculated by integrating the recorded averaged waveform. A cross-check using an optical power meter has confirmed that this method is accurate. We observe the energy of Alice's laser pulses with and without Eve's tampering laser. We have tested two ID300 short-pulse laser sources from ID~Quantique and one LP1550-SAD2 laser diode (LD) from Thorlabs. They are triggered by an external signal. ID300 contains a factory pre-set pulsed driver electronics and produces $50$--$70~\pico\second$ full width at half maximum~(FWHM) optical pulses, with their repetition rate controlled by our external electronic pulse generator (PG; Picosecond 12050). LP1550-SAD2's diode current is driven directly from the PG with pulse parameters set to obtain about $60~\pico\second$ FWHM optical pulses from the LD. The pulse repetition rate for all samples is $1~\mega\hertz$. The electronic pulse generator also acts as the external trigger of the oscilloscope as shown in~\cref{fig:scheme}.

\begin{figure*}
	\includegraphics[width=\textwidth]{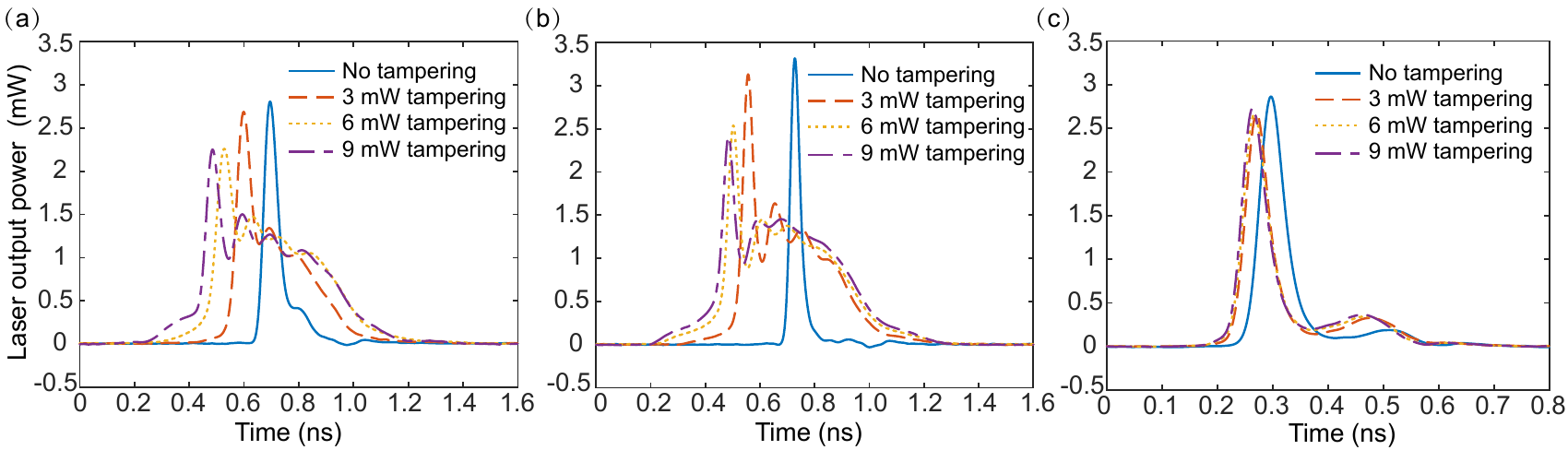}
	\caption{Averaged waveforms of laser pulses measured from (a)~ID300~sample~1, (b)~ID300~sample~2, and (c)~the laser diode LP1550-SAD2 from Thorlabs. Each oscillogram is an average over 2000 pulses.}
	\label{fig:ID300_wave}
\end{figure*}

\section{Experimental results}
\label{sec:result}

\begin{figure}
	\includegraphics[width=\columnwidth]{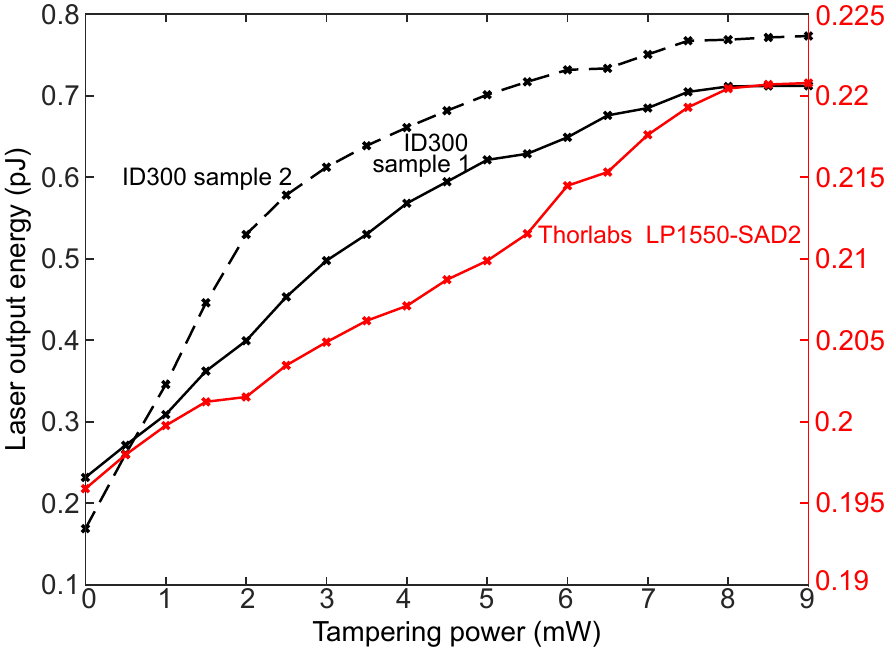}
	\caption{Average energy of Alice's output pulses as a function of Eve's tampering power for two samples of the laser ID300 from ID~Quantique (black curves) and the laser diode LP1550-SAD2 from Thorlabs (red curve). The energy of the pulse increases up to 3.07 times for ID300~sample~1, 4.57 times for ID300~sample~2, and 1.13 times for Thorlabs LP1550-SAD2.}
	\label{fig:ID300}
\end{figure} 

Both samples of ID300 exhibit controllability of their output power by Eve. We first measure the center wavelength of each laser with a spectrum analyser (Yokogawa AQ6370D). Then, in the experimental setup shown in~\cref{fig:scheme}, we dial the value of Alice's wavelength in Eve's laser. As a result, the output power of Alice's pulse suddenly increases. To obtain the maximum output power under Eve's control, we finely tune Eve's wavelength until the largest energy rise is observed, which is $1550.15~\nano\metre$ for sample~1 and $1550.44~\nano\metre$ for sample~2. This is the case we focus on. Additionally, we have noted that slightly different seed wavelengths result in different pulse shapes as shown in \cref{diff_wave}.

When we gradually increase the power of Eve's c.w.\ laser, the energy of Alice's emitted pulses also increases. This is shown in~\cref{fig:ID300_wave}~(a) and (b), which illustrates the waveforms of Alice's pulses for various tampering light powers. If we compare these results with the original waveform of Alice's pulses ({\it i.e.,}\ that in the absence of Eve's tampering laser), there are two main effects. First, as already mentioned, we see that the energy of the emitted optical pulses gets larger when we increase the tampering light power. Especially, Eve's injected light makes Alice's laser pulses wider with a much longer and higher tail as shown in~\cref{fig:ID300_wave}~(a) and (b). The tail contains more energy when higher power is injected into the diode. Second, under the laser seeding attack, the main peak of Alice's pulse shifts to be earlier. This is because the injected light triggers the stimulated emission that happens quicker than the spontaneous emission in Alice's laser diode. Thus, Alice's pulse reaches the peak power earlier and is followed by a tail with 2--4 secondary oscillations under the attack.

We have measured the energy of Alice's pulses for different tampering light powers. The results are shown in~\cref{fig:ID300} as black curves. In particular, we find that when there is no attack, this energy is $0.232~\pico\joule$~($0.169~\pico\joule$) for sample~1~(2). Then, we gradually increase the power of Eve's c.w.\ laser up to $9~\milli\watt$, and obtain that the output energy of Alice's laser rises up to $0.712~\pico\joule$~($0.773~\pico\joule$) for sample~1~(2). That is, the pulse energy increases 3.07~(4.57) times for sample~1~(2).

Under the same experimental procedure done with ID300, a similar effect is observed in the laser LP1550-SAD2. The wavelength of the injected c.w.\ light is set to the center wavelength of the laser diode first, then tuned slightly to $1551.32~\nano\metre$ where we observe the maximum increase in Alice's pulse energy. \Cref{fig:ID300_wave}~(c) shows the waveforms of Alice's pulses for the same tampering light powers as those in~\cref{fig:ID300_wave}~(a) and (b). Similarly to ID300 lasers, here the energy of the pulses increases with the tampering power as well. The rising edge of Alice's pulse also starts earlier in the presence of the attack. The increase of the pulse energy as a function of Eve's tampering power is shown in~\cref{fig:ID300} as a red curve. If there is no attack, the average energy of Alice's laser pulses is $0.196~\pico\joule$, while it reaches $0.221~\pico\joule$ when the tampering power is $9~\milli\watt$. That is, in this case the pulse energy increases 1.13 times.

We note that the commercial lasers under test in our experiment (ID300 and LP1550-SAD2) contain an internal optical isolator of the order of $30$--$40~\deci\bel$. Thus, a few $\milli\watt$ light that is applied in our experiment is first attenuated at the internal isolator of the laser, which means that only about $100~\nano\watt$ power actually reaches the laser cavity. This analysis indicates that an injection power in the order of $100~\nano\watt$ could be enough to control the intensity of Alice's pulses. Indeed, this value of injection power has been also confirmed recently by the experimental results shown in Ref.~\cite{pang2019}, in which Eve's injection power is in $100$--$160~\nano\watt$ range. We also note that a real QKD system may use a laser diode without the internal isolator, then the injection power used in our laser seeding attack may be reduced to the above level.

\section{Effect on the security of QKD}\label{sec:Key_rate}

Now we show theoretically how an unnoticed increase of the optical power emitted by a QKD transmitter, due to the attack described above, could seriously compromise the security of a QKD implementation. We assume that Alice's photon number statistics is Poissonian and is not influenced by our attack. The former may not necessarily be the case \cite{dynes2018}, and investigating the validity of the latter assumption could be the topic of a future study. Based on this assumption, we shall consider the case of decoy-state based QKD~\cite{hwang2003,wang2005a,lo2005}, which includes the most implemented QKD schemes today. We refer the reader to~\cref{Decoy} for further details about decoy-state based QKD. For simplicity, in our analysis we shall assume the asymptotic scenario where Alice sends Bob an infinite number of pulses, {\it i.e.,}\ we disregard statistical fluctuations due to finite size effects. Also, motivated by the experimental results presented in the previous section, we shall consider that Eve's attack increases all the intensities $\mu$ by the same factor $\kappa>1$. That is, we will assume that $\mu'=\kappa\mu$ for all $\mu$.

Next, we quantitatively evaluate the effect that a laser seeding attack has on the security of the standard decoy-state BB84 protocol and of MDI-QKD for a typical channel model. For concreteness, we will consider first the case of the standard decoy-state BB84 protocol with phase-randomized weak coherent pulses (WCPs); afterward, we will consider the case of MDI-QKD.

\subsection{Standard decoy-state BB84 protocol}

Regarding the standard decoy-state BB84 protocol, we evaluate the typical implementation where Alice and Bob use three different intensities, $\mu_s$, $\nu_1$ and $\nu_2$ that satisfy $\mu_s>\nu_1>\nu_2$, and they generate secret key only from those events where they employ the signal intensity $\mu_s$ in the $Z$ basis, while they use the $X$ basis events for parameter estimation. In the asymptotic limit of an infinite number of transmitted signals, the secret key rate can be lower bounded by~\cite{ma2005,gottesman2004}
\begin{eqnarray}\label{krate}
R_L = p_1^{\mu_s}Y_{1,L}^Z[1-H_2(e_{1,U}^X)] - f_eG_Z^{\mu_s}H_2(E_Z^{\mu_s}),\label{RateBB84}
\end{eqnarray}
where we assume the efficient version of this protocol~\cite{lo2005a}. In Eq.~(\ref{krate}), $Y_{1,L}^Z$ ($e_{1,U}^X$) denotes a lower (upper) bound on the single-photon yield $Y_1^Z$ (phase error rate $e_1^X$), the parameter $f_e$ is the error correction efficiency, $G_Z^{\mu_s}$ ($E_Z^{\mu_s}$) represents the overall experimentally observed gain (the overall experimentally observed QBER) when Alice send to Bob a WCP of intensity $\mu_s$ in the $Z$ basis, and $H_2(x)=-x\log_2{(x)}-(1-x)\log_2{(1-x)}$ is the binary Shannon entropy function.

To estimate $Y_{1,L}^Z$ and $e_{1,U}^X$ one can use analytical or numerical tools. Here we use the analytical method proposed in Ref.~\cite{ma2005}. In particular, we have that
\begin{eqnarray}
Y_{1,L}^Z&\geq&\frac{\mu_s}{\mu_s(\nu_1-\nu_2)-\nu_1^2+\nu_2^2}\bigg[G_Z^{\nu_1}e^{\nu_1}-\nonumber\\
&&G_Z^{\nu_2}e^{\nu_2}-\frac{\nu_1^2-\nu_2^2}{\mu_s^2}(G_Z^{\mu_s}e^{\mu_s}-Y_{0,L}^Z)\bigg]\label{Y1L},\\
e_{1,U}^X&\leq&\frac{E_X^{\nu_1}G_X^{\nu_1}e^{\nu_1}-E_X^{\nu_2}G_X^{\nu_2}e^{\nu_2}}{(\nu_1-\nu_2)Y_{1,L}^X}\label{e1U},
\end{eqnarray}
with $Y_{0,L}^Z$ being a lower bound on $Y_0^Z$ given by
\begin{equation}
Y_{0,L}^Z\geq\frac{\nu_1 G_Z^{\nu_2}e^{\nu_2}-\nu_2 G_Z^{\nu_1}e^{\nu_1}}{\nu_1-\nu_2}\label{Y0L},
\end{equation}
and where the parameter $Y_{1,L}^X$ represents a lower bound on $Y_1^X$. This last quantity can be obtained by using Eq.~(\ref{Y1L}) but now referred to the $X$ basis. 

In the presence of a laser seeding attack, Alice and Bob estimate $Y_{1,L}^Z$ and $e_{1,U}^X$ using Eqs.~(\ref{Y1L}) and (\ref{e1U}) but now with the experimentally observed quantities $G_\alpha^{\mu'}$ and $E_\alpha^{\mu'}$, with $\alpha\in\{Z,X\}$, $\mu'=\kappa\mu$ and $\mu\in\{\mu_s,\nu_1,\nu_2\}$ for a certain $\kappa$ that depends on the attack.

In our analysis we shall also evaluate an ultimate upper bound on the secret key rate. That is, this upper bound holds for 
any possible post-processing method that Alice and Bob could apply to their raw data. The only assumption here is that double click events are randomly assigned to single click events. For this, we use the technique introduced in Ref.~\cite{curty2009}. More precisely, the upper bound on the key rate is given by 
\begin{eqnarray}\label{Rupper}
R_U\leq\sum_{n\geq 1}r_{n}(1-\lambda_{\text{BSA}}^{n})I_{n}^{\text{ent}}(A;B),
\end{eqnarray}
where $r_{n}\approx e^{-\mu_s}\mu_s^{n}/n!$ is the probability that Alice sends Bob an $n$-photon state with the signal intensity, $\lambda_{\text{BSA}}^{n}$ is the maximum weight of separability among all the bipartite quantum states $\sigma_{AB}^n$ that are compatible with Alice and Bob's observables, and $I_{n}^{\text{ent}}(A;B)$ is the Shannon mutual information evaluated on the entanglement part of the state $\sigma_{AB}^n$ that has the maximum weight of separability. See Ref.~\cite{curty2009} and Appendix~\ref{ApDecoy} for further details.

\begin{figure}
	\includegraphics[width=\columnwidth]{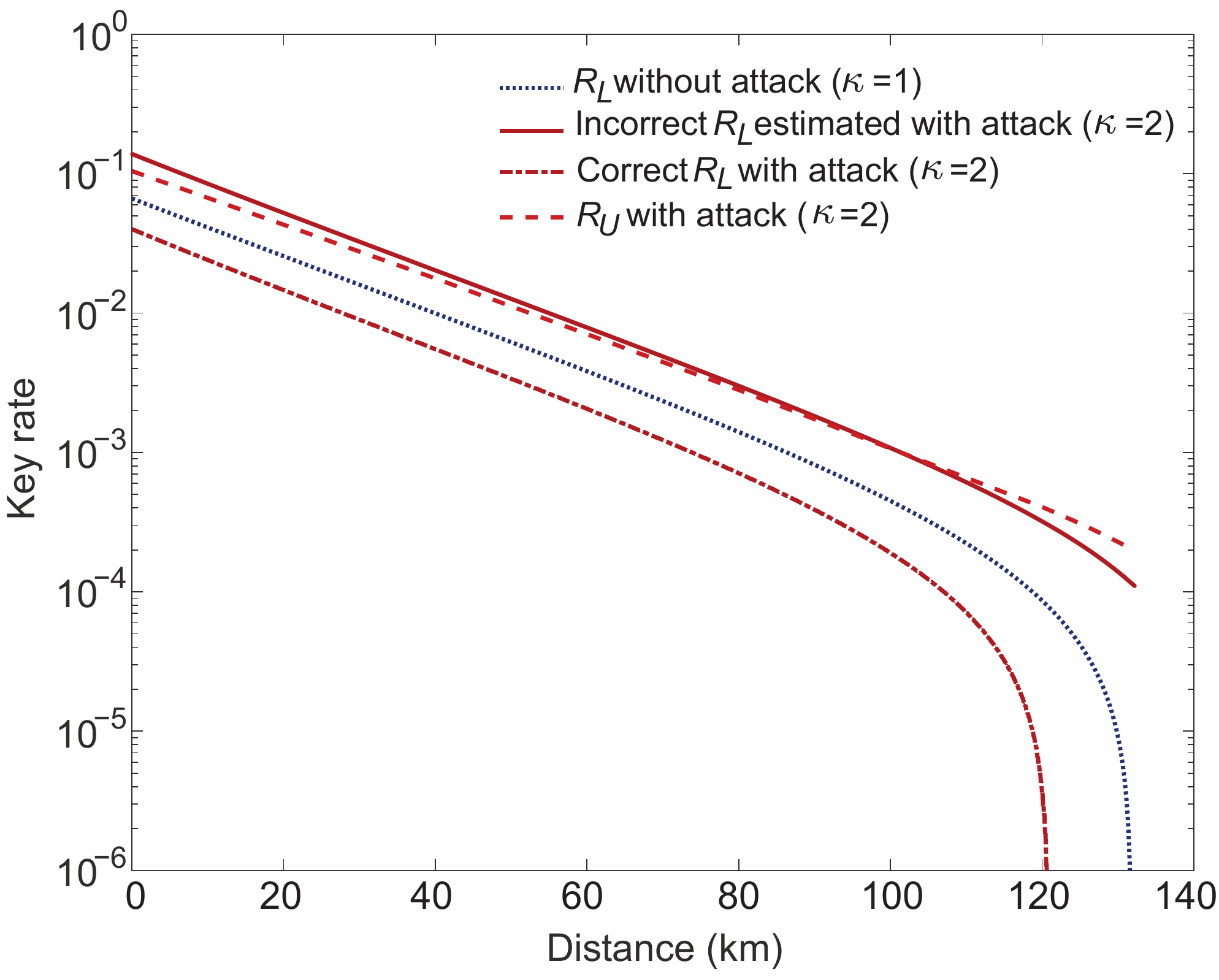}
	\caption{Lower ($R_L$) and upper ($R_U$) bounds on the secret key rate as a function of the distance for the standard decoy-state BB84 protocol for two different values of the multiplicative factor $\kappa=\{1,2\}$. The original intensity settings have been optimized previously for each distance. The parameters used in the simulations are given in~\cref{tab:experimental-parameters}.}\label{RateVSdistanceBB84}
\end{figure}

For simulation purposes we use the experimental parameters listed in~\cref{tab:experimental-parameters}. 
\begin{table}	
	\caption{Experimental parameters used in the simulations. The background rate and detection efficiency of the SPDs are taken from Ref.~\cite{comandar2016}.}\label{tab:experimental-parameters}
    \renewcommand*{\arraystretch}{1.15}
		\begin{tabular}[t]{lcc}
		\hline\hline
			Channel loss coefficient ($\deci\bel\per\kilo\meter$) & $\alpha$ & $0.2$ \\
			Background rate & $Y_0$ & $2.6\times10^{-5}$ \\
			Total misalignment error & $e_d$ &  1.5\% \\
			Detection efficiency of the SPDs & $\eta_D$ & 30\% \\
			Error correction efficiency & $f_e$ & 1.12 \\
		\hline\hline
		\end{tabular}
\end{table}
The resulting lower and upper bounds on the secret key rate are shown in Fig.~\ref{RateVSdistanceBB84}. 
The blue dotted line represents the lower bound $R_L$ given by Eq.~(\ref{RateBB84}) in the absence of the attack. Here, for each given value of the distance, we select the optimal values of the intensities $\mu_s$, $\nu_1$ and $\nu_2$ that maximize $R_L$. These optimized intensities are then fixed, and we use them to simulate the degradation of the security bounds due to Eve's laser seeding attack. 

More precisely, the red solid line in Fig.~\ref{RateVSdistanceBB84} shows the value of $R_L$ that Alice and Bob would estimate in the presence of the attack when $\kappa=2$. That is, as explained above, here Alice and Bob estimate the parameters $Y_{1,L}^Z$ and $e_{1,U}^X$ with the observed quantities $G_\alpha^{\mu'}$ and $E_\alpha^{\mu'}$, with $\alpha\in\{Z,X\}$, $\mu'=\kappa\mu$ and $\mu\in\{\mu_s,\nu_1,\nu_2\}$, together with the original intensities $\mu_s$, $\nu_1$ and $\nu_2$. The red dash-dotted line, on the other hand, illustrates the correct secure value of $R_L$ in the presence of the attack. That is, here $Y_{1,L}^Z$ and $e_{1,U}^X$ are estimated with the observed quantities $G_\alpha^{\mu'}$ and $E_\alpha^{\mu'}$, with $\alpha\in\{Z,X\}$, $\mu'=\kappa\mu$ and $\mu\in\{\mu_s,\nu_1,\nu_2\}$, together with the modified intensities $\mu'$. 

As we can see in Fig.~\ref{RateVSdistanceBB84}, the secure $R_L$ given by the red dash-dotted line is significantly below the $R_L$ actually estimated by Alice and Bob. That is, in the presence of the attack, the security proof introduced in Refs.~\cite{ma2005,gottesman2004} cannot guarantee the security of the secret key obtained by Alice and Bob. Finally, the red dashed line illustrates the upper bound $R_U$ given by Eq.~(\ref{Rupper}) in the presence of the attack. Remarkably, this upper bound is below the $R_L$ estimated by Alice and Bob for most of the distances, which demonstrates that the estimated secret key rate is actually insecure no matter what security proof is used.    

\begin{figure}
	\includegraphics[width=\columnwidth]{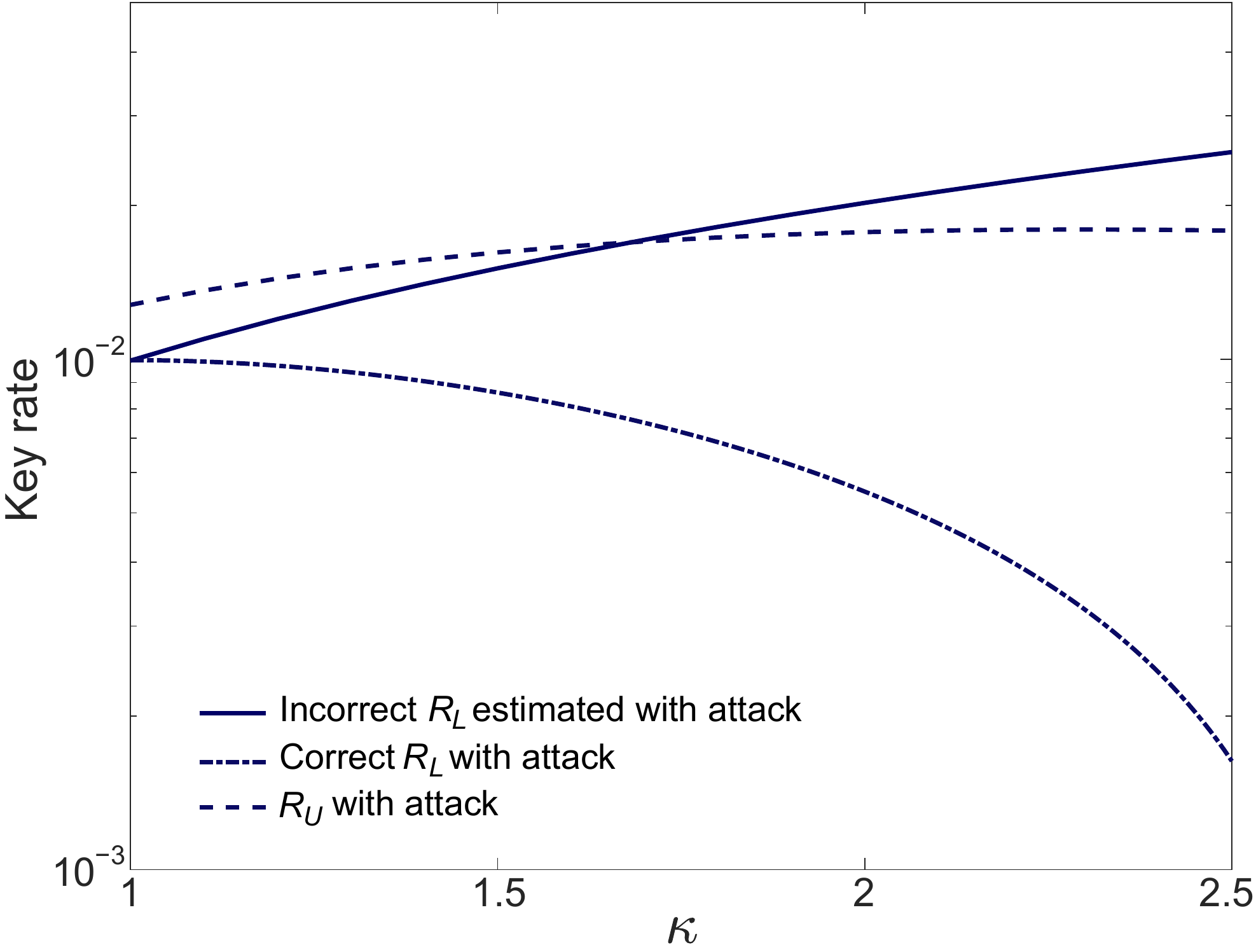}
	\caption{Lower ($R_L$) and upper ($R_U$) bounds on the secret key rate as a function of the parameter $\kappa$ for a fixed distance ($40$ km in this case), for decoy-state BB84 protocol. In these simulations, the original intensity settings have been optimized previously for the distance of 40 km assuming no attack. The parameters used in the simulations are given in~\cref{tab:experimental-parameters}.}\label{RateVSkappa}
\end{figure}

Finally, in Fig.~\ref{RateVSkappa} we show the effect that the multiplicative factor $\kappa$ has on the bounds on the secret key rate. For this, we now fix the transmission distance at a certain value, say $40~\kilo\meter$.  
In this case, Fig.~\ref{RateVSkappa} shows that the incorrect lower bound $R_L$ that Alice and Bob would estimate is always above its correct value whenever $\kappa>1$. This is remarkable because it means that in the presence of a laser seeding attack  Alice and Bob always overestimate their secret key rate above that provided by the security proof. Moreover, if $\kappa$ is large enough (around $1.7$ for the experimental parameters used in Fig.~\ref{RateVSkappa}), it turns out that the upper bound $R_U$ is below the estimated secret key rate, which confirms that there is no security proof which can make the estimated secret key rate secure. 

We remark that in practice Eve might need to throttle the key rate to roughly the original expected level in the absence of the attack. Indeed a human operator of QKD equipment may suspect something abnormal is happening on if the key generation rate rises well above the expected level (blue dotted line in \cref{RateVSdistanceBB84}). To reduce the rate, Eve can simply introduce additional optical attenuation in the channel.

\subsection{MDI-QKD}

Next we consider the case of MDI-QKD with WCPs~\cite{lo2012}. Similar to the previous example, we shall assume that each of Alice and Bob use three different intensities, $\mu_s$, $\nu_1$ and $\nu_2$ that satisfy $\mu_s>\nu_1>\nu_2$, and they generate secret key from those events encoded with the signal intensity in the $Z$ basis, while they use the $X$ basis events for parameter estimation. In the asymptotic limit of an infinite number of transmitted signals (and assuming for simplicity a sifting factor $\approx 1$), the secret key rate is lower bounded by~\cite{lo2012}
\begin{eqnarray}
R_L = p_{11}^{\mu_s\mu_s}Y_{11,L}^{Z}[1-H_2(e_{11,U}^{X})] - f_eG_{Z}^{\mu_s\mu_s}H_2(E_{Z}^{\mu_s\mu_s}),\ \ \ \ \label{RateMDI}
\end{eqnarray}
where $p_{11}^{\mu_s\mu_s}$ is the probability that both Alice and Bob emit a single-photon pulse in the $Z$ basis when they both use the signal intensity setting $\mu_s$, $Y_{11,L}^{Z}$ is a lower bound on the yield associated to these single-photon events, $e_{11,U}^{X}$ is an upper bound on the phase error rate of these single-photon pulses, $f_e$ is again the error correction efficiency, $G_{Z}^{\mu_s\mu_s}$ and $E_{Z}^{\mu_s\mu_s}$ are the gain and the QBER when both Alice and Bob send to the relay Charles WCPs of intensity $\mu_s$ in the $Z$ basis, and $H_2(x)$ is the binary Shannon entropy function defined previously.

To evaluate Eq.~(\ref{RateMDI}), Alice and Bob need to calculate the parameters $Y_{11,L}^{Z}$ and $e_{11,U}^{X}$ based on the experimentally available data $G_{\alpha}^{\zeta\omega}$ and $E_{\alpha}^{\zeta\omega}$, with $\alpha\in\{Z,X\}$ and $\zeta, \omega\in\{\mu_s,\nu_1,\nu_2\}$, and their knowledge on the probability distribution $p_{nm}^{\zeta\omega}$ with $n,m\in\mathbb{N}$, where $\mathbb{N}$ is the set of the non-negative integers. Again, this estimation can be done analytically or numerically, and for simplicity here we use the analytical approach introduced in Ref.~\cite{xu2014a}. For completeness, below we include the expressions for $Y_{11,L}^{Z}$ and $e_{11,U}^{X}$:
\begin{eqnarray}
	Y_{11,L}^{Z}&\geq&\frac{1}{(\mu_s-\nu_2)^2(\nu_1-\nu_2)^2(\mu_s-\nu_1)^2}\nonumber\\
	&&\times[(\mu_s^2-\nu_2^2)(\mu_s-\nu_2)(G^{\nu_1\nu_1}_{Z}e^{2\nu_1}+G^{\nu_2\nu_2}_{Z}e^{2\nu_2}\nonumber\\
	&&-G^{\nu_1\nu_2}_{Z}e^{\nu_1+\nu_2})-(\nu_1^2-\nu_2^2)(\nu_1-\nu_2)(G_{Z}^{\mu_s\mu_s}e^{2\mu_s}\nonumber\\
	&&+G^{\nu_2\nu_2}_{Z}e^{2\nu_2}-G^{\mu_s\nu_2}_{Z}e^{\mu_s+\nu_2}-G^{\nu_2\mu_s}_{Z}e^{\nu_2+\mu_s})], \label{uot}
\end{eqnarray}
and
\begin{eqnarray}
	e_{11,U}^{X}&\leq&\frac{1}{(\nu_1-\nu_2)^2Y_{11,L}^{X}}(e^{2\nu_1}G_X^{\nu_1\nu_1}E_X^{\nu_1\nu_1}+e^{2\nu_2}G_X^{\nu_2\nu_2}\nonumber\\
	&&\times{}E^{\nu_2\nu_2}_X-e^{\nu_1+\nu_2}G_X^{\nu_1\nu_2}E_X^{\nu_1\nu_2}-e^{\nu_2+\nu_1}G_X^{\nu_2\nu_1}\nonumber\\
	&&\times{}E_X^{\nu_2\nu_1}), \label{eux}
\end{eqnarray}
where $Y_{11,L}^{X}$ represents a lower bound on the yield associated to those single-photon events emitted by Alice and Bob in the $X$ basis. This last quantity can be estimated using Eq.~(\ref{uot}) but now referred to the $X$ basis. 

To evaluate $R_L$ in the presence of a laser seeding attack we follow a methodology similar to that used in the previous subsection, and we omit it here for simplicity. 

Also, to evaluate an upper bound $R_U$ on the secret key rate, we extend the technique introduced in Ref.~\cite{curty2009} to the case of MDI-QKD. Here, for simplicity, we consider that Alice and Bob only distill secret key from nonpositive partial transposed entangled states~\cite{peres1996,horodecki1996}, {\it i.e.,}\ we disregard the key material which could be obtained from positive partial transposed entangled states~\cite{horodecki2005}. We refer the reader to Appendix~\ref{ApMDI} for further details about the upper bound $R_U$. 

\begin{figure}
	\includegraphics[width=\columnwidth]{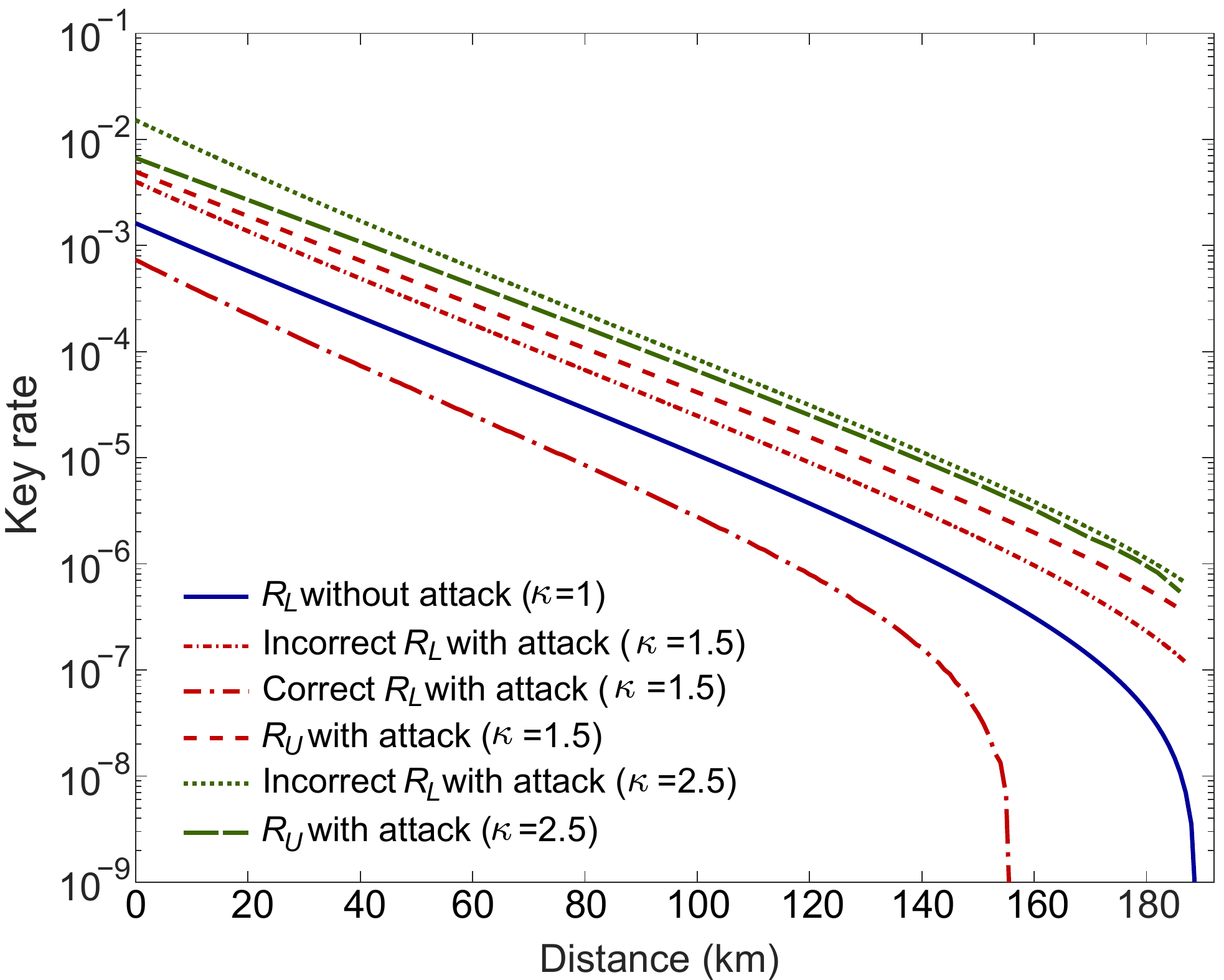}
	\caption{Lower ($R_L$) and upper ($R_U$) bounds on the secret key rate as a function of the distance for MDI-QKD with WCPs for three different values of the multiplicative factor $\kappa=\{1,1.5,2.5\}$. The correct value of $R_L$ in the presence of the attack is zero when $\kappa=2.5$. This shows that Alice and Bob significantly overestimate the secret key rate in the presence of the attack. The original intensity settings have been optimized previously for each distance for the case where there is no attack. The parameters used in the simulations are given in~\cref{tab:experimental-parameters}.}\label{RateVSdistance_mdiQKD}
\end{figure}

For simulation purposes, we use again the experimental parameters given in~\cref{tab:experimental-parameters}. For simplicity, we assume that Eve performs a symmetric attack in which she injects light of the same intensity into both Alice's and Bob's transmitter devices, which moreover we assume are identical. The resulting lower and upper bounds on the secret key rate are shown in Fig.~\ref{RateVSdistance_mdiQKD}.  
For this example we consider three possible values for the multiplicative factor $\kappa=\{1,1.5,2.5\}$. The case $\kappa=1$ corresponds to the scenario without attack. The results are analogous to those illustrated in Fig.~\ref{RateVSdistanceBB84}. In particular, the incorrect value of $R_L$ that Alice and Bob would estimate in the presence of the attack is well above the correct value of $R_L$ delivered by a proper application of the security proof ({\it i.e.,}\ for the case where one considers the correct values of the output intensities modified by the attack). This is particularly critical for the case where $\kappa=2.5$, as the security proof provides no secure key rate in this scenario while Alice and Bob would incorrectly estimate a relatively high value for $R_L$. Also, in this case, the upper bound $R_U$ is below the estimated $R_L$ for all distances (see Fig.~\ref{RateVSdistance_mdiQKD}).

\section{Discussion and countermeasure}\label{sec:discussion}

In this laser seeding attack, the isolation present in a real QKD system may significantly affect Eve's injection power. Thus, we should analyse this effect in detail. The first factor that contributes to such isolation is the presence of an attenuator to attenuate Alice's signals to the single-photon level. If we assume that 
the power of Alice's laser is similar to the laser we tested, the required attenuation would be in the order of $60~\deci\bel$ to obtain single-photon-level pulses. This means that Eve's initial injection laser (before going through the attenuator) should be in the order of $100~\milli\watt$ (assuming that there is no internal isolator in the laser) such that about $100~\nano\watt$ power can enter the laser cavity. This value is reasonable and can be safely transmitted through optical fiber, which confirms that the laser seeding attack is practical.

Furthermore, we note that the attenuation provided by optical attenuators can be decreased via a laser damage attack~\cite{huang2019b}. Specifically, Eve can illuminate Alice's attenuator with a c.w.\ laser with power of several watts. The experimental results reported in~\cite{huang2019b} show that it is possible to permanently decrease the attenuation by more than $10~\deci\bel$ by the c.w.\ laser. Importantly, this can be done such that no connector or other components in the experiment are damaged. The attenuator is the only component that responds. Therefore, if Eve applies first the laser damage attack against the attenuators to decrease their attenuation, then the injection power of the laser seeding attack could be even lower than $100~\milli\watt$. This strategy of combination attacks makes the laser seeding attack easier to implement thanks to the laser damage attack.

The second factor that could contribute to have more isolation is to include an external isolator. The isolator indeed makes Eve's attack more difficult. However, according to the working mechanism of an optical isolator, the isolation of the backward injection light is due to the polarization rotation inside the isolator, after which the rotated light is extinguished. The rotation is realized by a magneto-optic effect. It is notable that the magnets used in isolators are temperature-dependent~\cite{vojna2019}. That is, the higher temperature, the smaller rotation. Thus, the temperature is an important factor in practice to determine the real isolation value. From Eve's point of view, she may somehow hack the isolator by increasing the temperature. The quantitative study of the dependence between the optical isolation provided by an optical isolator and the temperature that Eve can achieve is beyond the scope of this paper, but we've studied this topic in another manuscript~\cite{ponosova2019}.

It is clear that for a given power of Eve's injected light, the more effective isolation the users'  transmitters have, the smaller the value of the multiplicative factor $\kappa$ will be, and thus also the effectiveness of the attack. For example, according to Fig.~\ref{fig:ID300}, if the power of Eve's injected light is say $10~\watt$, then an effective isolation $ > 80~\deci\bel$ would result in a multiplicative factor $\kappa<2$ for ID300 sample 2. Importantly, however, as we have seen in~\cref{RateVSkappa}, whenever $\kappa>1$ (which in principle might happen even for very high isolation), Alice and Bob might always overestimate their secret key rate, unless, of course, they modify their security analysis to properly incorporate the effect of the laser seeding attack.

For this, for instance, Alice and Bob could first bound the power of Eve's injected light to a reasonable value, as done for example in Refs.~\cite{lucamarini2015,tamaki2016,wang2018,huang2019b}. With this assumption in place, and for a given value of the isolation of their transmitters, as well as the behaviour of their laser sources, Alice and Bob could in principle upper bound the maximum value, $\kappa_{\rm max}$, that the parameter $\kappa$ can take. In so doing, and for given observed experimental data ({\it i.e.,}\ gains and error rates associated to different values of the intensity settings), they could simply minimize their secret key rate by taking into account that now the intensities of the emitted light pulses might lay in an interval $[\mu,\kappa_{\rm max}\mu]$, where $\mu$ is the value of the original intensity setting. This way Alice and Bob consider the worst-case scenario and can guarantee that the resulting secret key rate is indeed secure.  

Another alternative for Alice to determine the parameter $\kappa_{\rm max}$ might be to use an incoming-light monitor to detect the injection light. The main drawback of this approach is, however, that the classical monitor that detects the injected light is not a reliable device. For example, in Ref.~\cite{sajeed2015}, it has been shown that the classical monitor can be bypassed by Eve's pulses with high repetition rate, and thus the classical monitor cannot correctly quantify the amount of injected light. This is due to the limited bandwidth of the classical monitor. Furthermore, the classical monitor may even be damaged by Eve's light~\cite{makarov2016}. According to the experimental results in Ref.~\cite{makarov2016}, the classical monitor is the first component in Alice that is damaged by Eve's laser. Therefore, the classical detector also may not be a reliable countermeasure to prevent Eve's injection of light.
 
In practice, it is important to note as well that Eve could in principle combine the laser seeding attack with various attacks to enhance her hacking capability, for example, with the laser damage attack~\cite{makarov2016,huang2019b} as mentioned above, with the THA analysed in Refs.~\cite{vakhitov2001,gisin2006,lucamarini2015,tamaki2016,wang2018,pereira2019}, and/or with the recently introduced injection-locking attack~\cite{pang2019}. For instance, Eve could employ the fact that the laser seeding can be affected in real time by the state of Alice's modulator, changing the laser wavelength depending on the modulator setting \cite{pang2019} and/or modulating the intensity multiplication factor $\kappa$. Besides using her injected light to modify the internal functioning of the transmitter (as done in the laser seeding attack), Eve could also simultaneously perform a THA and measure the back-reflected light to obtain information about the transmitter's settings for each emitted light pulse. This means that to properly evaluate the security of a QKD system, one should probably combine the techniques described in the previous paragraphs with the security analysis introduced in Refs.~\cite{lucamarini2015,tamaki2016,wang2018,pereira2019}.
 
\section{Conclusion}
\label{sec:conclusion}

This study has experimentally demonstrated that the laser seeding attack is able to increase the intensity of the light emitted by the laser diode used in a QKD system, breaking the fundamental assumption about the mean photon number of a QKD protocol. Moreover, we have shown theoretically that such increase of the intensity might seriously compromise the security of QKD implementations. For this, we have considered two prominent examples: the standard decoy-state BB84 protocol and MDI-QKD, both implemented with phase-randomized WCPs. In both cases, we have demonstrated that, in the presence of the attack, the legitimate users of the system might significantly overestimate the secret key rate provided by proper security proofs, even well above known upper bounds. This theoretical security analysis can be applied to any attack that increases the intensity of the emitted pulses. For instance, a laser damage attack against the optical attenuators also shows that Eve can increase the intensity of Alice's pulses by decreasing the attenuation provided by the attenuators~\cite{huang2019b}.

Although MDI-QKD is immune to all detector side-channel attacks, our work shows Eve's capability of hacking the source of a QKD system and highlights that further research is needed to protect the system against source side-channel attacks. Moreover, we remark that the laser seeding attack may compromise as well the security of other quantum decoy-state based cryptographic systems beyond QKD, like, for instance, various two-party protocols with practical signals~\cite{wehner2010}, quantum digital signatures~\cite{yin2016b, roberts2017}, and blind quantum computing~\cite{xu2015c,zhao2017}.

While preparing this Article for publication, we have learned of another laser seeding experiment that changes the wavelength of Alice's laser rather than its intensity~\cite{pang2019}.

\def\bibsection{\medskip\begin{center}\rule{0.5\columnwidth}{.8pt}\end{center}\medskip} 

%

\acknowledgments

We thank Koji Azuma for very useful discussions. This work was funded by the Galician Regional Government (program atlanTTic), the Spanish Ministry of Economy and Competitiveness (MINECO), Fondo Europeo de Desarrollo Regional (FEDER; grants TEC2014-54898-R and TEC2017-88243-R), the European Union's Horizon 2020 program under the Marie Sk\l{}odowska-Curie project QCALL (grant 675662), the National Natural Science Foundation of China (grants 11674397, 61601476, 61632021, and 61901483), the National Key Research and Development Program of China (Grant 2019QY0702), NSERC of Canada (programs Discovery and CryptoWorks21), CFI, MRIS of Ontario, and the Ministry of Education and Science of Russia (NTI center for quantum communications). A.H.\ was supported by China Scholarship Council. {\'A}.N.\ was supported from the Spanish Ministry of Education (FPU scholarship). P.C.\ was supported by the Thai DPST scholarship.

{\em Author contributions:} A.H.\ performed the experimental testing with the assistance of S.-H.S.\ and P.C. {\'A}.N.\ and M.C.\ performed the key rate analysis. V.M.\ and M.C.\ supervised the study. A.H.,\ {\'A}.N.,\ and M.C.\ wrote the article with input from all authors.

\clearpage
\appendix

\section{Laser seeding by different wavelengths}
\label{diff_wave}

\begin{figure}
	\includegraphics{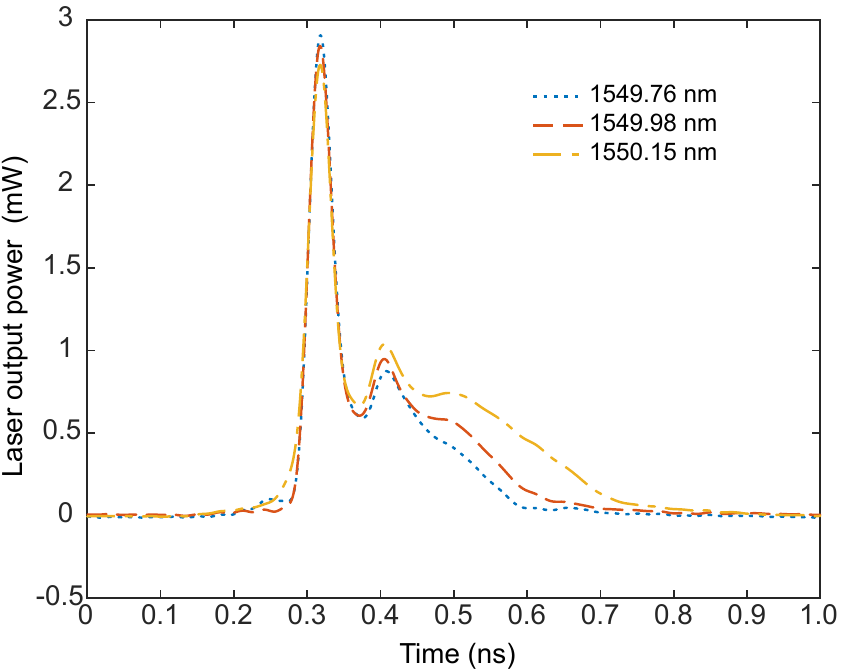}
	\caption{Averaged waveforms of the laser pulses measured from ID300 sample 1 with $2~\milli\watt$ tampering power for different wavelengths. Each oscillogram is an average over 2000 pulses.}
	\label{fig:ID300_diff_wave}
\end{figure}

In the laser seeding attack, we pick the wavelength of the injected light to obtain the maximum energy of Alice's optical pulses. At this wavelength, we observe the increased energy and the longer tail, as shown in~\cref{fig:ID300_wave}. Moreover, we have tested the injected light with slightly different wavelengths that are still in the wavelength range of the laser diode from the high-speed oscilloscope, see \cref{fig:ID300_diff_wave}. Sample 1 of ID300 with $1~\nano\meter$ linewidth is shown as an example. When $2~\milli\watt$ power is injected into the laser, different wavelengths result in different waveforms. At $1550.15~\nano\meter$, Alice's pulse has the highest energy but relatively lower peak power. When the wavelength is slightly off the center wavelength, at $1549.98~\nano\meter$, the peak power becomes higher, however the tail is lower. This trend continues when the wavelength is shifted further to $1549.76~\nano\meter$.

\section{Decoy-state QKD protocol}
\label{Decoy}

In decoy-state QKD, the transmitter emits quantum states that are diagonal in the Fock basis, and whose mean photon number is selected at random, within a predefined set of possible values, for each output signal. These states are typically generated with an attenuated laser diode emitting phase-randomized weak coherent pulses (WCPs) in combination with a variable attenuator to set the intensity of each individual light pulse.

In particular, let $Y_n^{\alpha}$ ($e_n^{\alpha}$) denote the $n$-photon yield (error rate) in the polarization basis $\alpha\in\{Z,X\}$. That is, $Y_n^{\alpha}$ ($e_n^{\alpha}$) represents the probability that an $n$-photon state prepared in the $\alpha$ basis generates a detection click (a detection click associated to an error in the $\alpha$ basis) at Bob's side. For each intensity setting $\mu$, these quantities are related to the overall experimentally observed gain, $G^{\mu}_{\alpha}$, and to the overall experimentally observed error rate, $E^{\mu}_{\alpha}$, in the $\alpha$ basis as follows:
\begin{eqnarray}\label{RelGood}
\begin{aligned}
	G^{\mu}_{\alpha}=\sum_n p_{n}^{\mu}Y_n^{\alpha},\\
	E^{\mu}_{\alpha}=\frac{1}{G^{\mu}_{\alpha}}\sum_n p_{n}^{\mu}e_n^{\alpha}Y_n^{\alpha},
\end{aligned}
\end{eqnarray}
where $p_{n}^{\mu}$ denotes the probability that Alice emits an $n$-photon state when she selects the intensity setting $\mu$. In the case of WCPs, these probabilities follow a Poissonian distribution, $p_{n}^{\mu}=e^{-\mu}\mu^n/n!$, that only depends on the mean photon number $\mu$. That is, $G^{\mu}_{\alpha}$ ($E^{\mu}_{\alpha}$) represents the probability that a WCP of intensity $\mu$ prepared in the $\alpha$ basis generates a detection click (a detection click associated to an error in the $\alpha$ basis) at Bob's side.

Importantly, Eq.~(\ref{RelGood}) relates the observed quantities $G^{\mu}_{\alpha}$ and $E^{\mu}_{\alpha}$ with the unknown parameters $Y_n^{\alpha}$ and $e_n^{\alpha}$ through the {\it known} probabilities $p_{n}^{\mu}$. This means, in particular, that by solving the set of linear equations given by Eq.~(\ref{RelGood}) for different values of $\mu$ one can obtain tight bounds on the relevant parameters $Y_1^Z$ and $e_1^X$ which are required to determine the resulting secret key rate. 

Now suppose that Eve performs a laser seeding attack that increases the output intensity of the emitted pulses from $\mu$ to say $\mu'$. In this scenario, Alice and Bob, who are unaware of the attack, would use the experimentally observed quantities $G^{\mu'}_{\alpha}$ and $E^{\mu'}_{\alpha}$, which depend on the modified mean photon number $\mu'$, together with the original (but now {\it erroneous}) probabilities $p_{n}^{\mu}$ that depend on the original intensity $\mu$, to estimate the parameters $Y_1^Z$ and $e_1^X$. That is, if Eve implements a laser seeding attack, Alice and Bob would use the following set of linear equations to estimate $Y_1^Z$ and $e_1^X$:
\begin{eqnarray}
\begin{aligned}
G^{\mu'}_{\alpha}=\sum_n p_{n}^{\mu}Y_n^{\alpha},\label{RelFake}\\
E^{\mu'}_{\alpha}=\frac{1}{G^{\mu'}_{\alpha}}\sum_n p_{n}^{\mu}e_n^{\alpha}Y_n^{\alpha}.
\end{aligned}
\end{eqnarray}
In so doing, the bounds obtained for $Y_1^Z$ and $e_1^X$ by solving Eq.~(\ref{RelFake}) are not guaranteed to be correct bounds for the single-photon yield in the $Z$ basis nor for the phase error rate. Indeed, the correct bounds for these two quantities satisfy Eq.~(\ref{RelGood}) after substituting $\mu$ with $\mu'$.

\section{Upper bound \bm{$R_U$} for decoy-state QKD}\label{ApDecoy}

Here we briefly summarize the technique introduced in Ref.~\cite{curty2009} to derive an upper bound on the secret key rate for a decoy-state QKD protocol. It basically consists in finding the best separable approximation (BSA)~\cite{lewenstein1998} among all bipartite quantum states that are compatible with the measurement results observed by Alice and Bob in an execution of the protocol. That is, these are the states that Alice and Bob could have shared in a virtual entanglement protocol that is equivalent to the actual protocol. For simplicity, Ref.~\cite{curty2009} considers a decoy-state protocol where Alice and Bob use an infinite number of decoy settings. Note, however, that in the asymptotic limit where Alice sends Bob an infinite number of signals, an upper bound on the secret key rate for this protocol applies as well to a protocol using a finite number of decoy settings. We follow the same procedure here.

In particular, let $S^{n}$ denote the set of all bipartite quantum states, $\sigma_{AB}^n$, which are compatible with Alice and Bob's measurement results in a virtual entanglement protocol that is equivalent to the actual protocol when Alice sends Bob an $n$-photon signal. That is, this set is defined as 
\begin{eqnarray}
S^{n}=\{\sigma_{AB}^n|\Tr[A_k\otimes B_j \sigma_{AB}^n]=p_{kj}^{n}\ \ \forall k,j\},
\end{eqnarray}
where $\{A_k\}_k$ and $\{B_j\}_j$ are the measurement operators of Alice and Bob in the virtual entanglement protocol, and $p_{kj}^n$ represent the measured statistics associated to the $n$-photon signals emitted by Alice. Since we assume that Alice uses an infinite number of decoy intensities, we consider that she can estimate these probabilities precisely.

The states $\sigma_{AB}^n\in{}S^{n}$ can always be expressed as a convex sum of one separable state, $\sigma_{\text{sep}}^n$, and one entangled state, $\rho_{\text{ent}}^n$, as follows
\begin{equation} \label{bsap}
\sigma_{AB}^n=\lambda_n \sigma_{\text{sep}}^n+(1-\lambda_n) \rho_{\text{ent}}^n,
\end{equation}
for some real parameter $\lambda_n\in[0,1]$. Then, the BSA of the states in $S^{n}$ corresponds to that state with the maximum value of the parameter $\lambda_n$, which we shall denote by $\lambda_{\text{BSA}}^n$. That is, for every $n$, we want to find the parameter
\begin{eqnarray}
\lambda_{\text{BSA}}^n=\max[\lambda_n|\sigma_{AB}^n\in{}S^{n}],\label{lambdaBSA}
\end{eqnarray}
as well as the corresponding entangled state $\rho_{\text{ent}}^n$ for the BSA. 

In standard decoy-state QKD with four sending states, Alice's measurement operators $\{A_k\}_k$ can be described by a projective measurement in a four-dimensional Hilbert space, {\it i.e.,}\ $A_k=\dyad{k}{k}$ with $k\in\{1,2,3,4\}$. Each operator $A_k$ is associated with Alice sending one of the four possible polarization states of the BB84 protocol. On Bob's side, his measurement operators $\{B_j\}_j$ correspond to a positive-operator valued measurement (POVM) with the following elements
\begin{eqnarray}
B_0=\frac{1}{2}\dyad{0}{0}, &\text{ }&B_1=\frac{1}{2}\dyad{1}{1},\nonumber\\
B_{\pm}=\frac{1}{2}\dyad{\pm}{\pm}, &\text{ }&B_{vac}=\dyad{vac}{vac},
\end{eqnarray}
where $\ket{\pm}=\frac{1}{\sqrt{2}}(\ket{0}\pm\ket{1})$, and $\ket{vac}$ is the vacuum state. As already mentioned in the main text, here we implicitly assume that double click events are randomly assigned by Bob to single click events.

In addition, we have that in a prepare\&{}measure QKD scheme the reduced density matrix of Alice, $\rho^{n}_{A}=\Tr_B(\sigma^{n}_{AB})$, is fixed by her state preparation process. In the scenario considered, it turns out that $\rho^{n}_{A}$ can be written as~\cite{curty2009}
\[
\rho^{n}_{A}=\frac{1}{4}
\begin{bmatrix}
1 		 & 0 		  & 2^{-n/2} & 2^{-n/2} \\
0 	     & 1 		  & 2^{-n/2} & (-1)^n2^{-n/2} \\
2^{-n/2} & 2^{-n/2} & 1 	     & 0 \\
2^{-n/2} & (-1)^n2^{-n/2} & 0 	     & 1
\end{bmatrix}.
\]
\begin{eqnarray}
\label{ReducedMatrix}
\end{eqnarray}

Putting all the conditions together, one can obtain the parameter $\lambda_{\text{BSA}}^n$ and the corresponding entangled state $\rho_{\text{ent}}^n$, for each $n$, by solving the following semidefinite program (SDP)~\cite{curty2009}
\begin{eqnarray}\label{sdp_decoy}
\min &\text{ }& 1-\Tr(\sigma^{n}_{\text{sep}}(\bm x)),\nonumber\\
\text{s.t.} && \sigma^{n}_{AB}(\bm x)\geq 0,\nonumber\\
&&\Tr[\sigma^{n}_{AB}(\bm x)]=1,\nonumber\\
&&\Tr[A_k\otimes B_j\sigma^{n}_{AB}(\bm x)]=p_{kj}^{n}, \forall k,j,\nonumber\\
&&\Tr_B(\sigma^{n}_{AB}(\bm x))=\rho^{n}_{A}, \nonumber\\
&& \sigma^{n}_{\text{sep}}(\bm x) \geq 0, \nonumber\\
&& \sigma^{n,\Gamma}_{\text{sep}}(\bm x) \geq 0, \nonumber\\
&& \sigma^{n}_{AB}(\bm x)-\sigma^{n}_{\text{sep}}(\bm x) \geq 0, 
\end{eqnarray}
where the vector $\bm x$ is used to parametrize the density operators and $\Gamma$ denotes partial transposition of one of the subsystems. Note that in Eq.~(\ref{sdp_decoy}) the state $\sigma^{n}_{\text{sep}}(\bm x)$ represents an unnormalized state, {\it i.e.,}\ if we compare this state with that given in Eq.~(\ref{bsap}) we have that $\sigma^{n}_{\text{sep}}(\bm x)=\lambda_n \sigma_{\text{sep}}^n$. 

From the optimal solution, $\bm x_{sol}$, of the SDP above we have that 
\begin{eqnarray}
\lambda_{\text{BSA}}^n&=&\Tr(\sigma^{n}_{\text{sep}}(\bm x_{sol})), \nonumber \\
\rho_\text{ent}^{n}&=&\frac{\sigma^{n}_{AB}(\bm x_{sol})-\sigma^{n}_{\text{sep}}(\bm x_{sol})}{1-\lambda_{\text{BSA}}^{n}}.
\end{eqnarray}

The upper bound on the secret key rate can then be written as~\cite{moroder2006,curty2009}
\begin{eqnarray}\label{up}
R_U\leq\sum_{n\geq 1}r_{n}(1-\lambda_{\text{BSA}}^{n})I_{n}^{\text{ent}}(A;B),
\end{eqnarray}
where $r_{n}\approx e^{-\mu}\mu^{n}/n!$ is the probability that Alice sends Bob an $n$-photon state, where $\mu$ is the mean photon number of the signal, and $I_{n}^{\text{ent}}(A;B)$ is the Shannon mutual information evaluated on $q_{kj}^{n}=\text{Tr}(A_k\otimes B_j \rho_\text{ent}^{n})$. Note that to calculate Eq.~(\ref{up}) it is typically sufficient to consider only a finite number of terms in the summation, because of the limit imposed by the unambiguous state discrimination attack. See~Ref.~\cite{curty2009} for further details.

\section{Upper bound \bm{$R_U$} for MDI-QKD}\label{ApMDI}

Here we extend the results in Ref.~\cite{curty2009} to the MDI-QKD framework to calculate an upper bound on the secure key rate coming, for simplicity, from nonpositive partial transposed entangled states~\cite{peres1996,horodecki1996}. Like in~Ref.~\cite{curty2009}, we consider for simplicity that Alice and Bob use an infinite number of decoy settings (see also Appendix~\ref{ApDecoy}).

In MDI-QKD, both Alice and Bob are transmitters while, in the middle, an untrusted third party Charles is supposed to perform a Bell state measurement on the incoming signals and publicity announce the result. Let $c\in S_{\text{an}}$ denote Charles' announcement, where $S_{\text{an}}$ is the set of all possible announcements. This set includes the possible Bell states that Charles can obtain with his measurement as well as the inconclusive event. For each announcement $c$, we will denote the set of bipartite quantum states, $\sigma_{AB,c}^{nm}$, that Alice and Bob could have shared in an equivalent virtual entanglement protocol (given that in the actual protocol they sent $n$ and $m$ photons to Charles, respectively) as $S^{nm}_c$. That is, $S^{nm}_c$ contains all the bipartite quantum states $\sigma_{AB,c}^{nm}$ that are compatible with Alice and Bob's measurement outcomes in the equivalent virtual entanglement protocol, 
\begin{eqnarray}
S^{nm}_c=\{\sigma_{AB,c}^{nm}|\Tr[A_k\otimes B_j \sigma_{AB,c}^{nm}]=p_{kj}^{nmc}\ \ \forall k,j\},\ \ \ \ \ 
\end{eqnarray}
where $\{A_k\}_k$ and $\{B_j\}_j$ are the measurement operators of Alice and Bob in the virtual entanglement protocol, and $p_{kj}^{nmc}$ represent the measured statistics associated to Charles' announcement $c$ when Alice (Bob) sends him an $n$-photon ($m$-photon) signal. In the same way as in Appendix~\ref{ApDecoy}, here it is assumed that Alice and Bob can estimate the probabilities $p_{kj}^{nmc}$ precisely because they use an infinite number of decoy intensities.

Similar to the case of the standard decoy-state BB84 protocol considered previously, we have that the states $\sigma_{AB,c}^{nm}\in{}S^{nm}_c$ can always be decomposed as the convex sum of a separable state, $\sigma^{nm}_{\text{sep},c}$, and an entangled state, $\rho^{nm}_{\text{ent},c}$, as follows
\begin{equation}
\sigma_{AB,c}^{nm}=\lambda_{nm}^c\sigma^{nm}_{\text{sep},c}+(1-\lambda_{nm}^c)\rho^{nm}_{\text{ent},c},
\end{equation}
for some real parameter $\lambda_{nm}^c\in[0,1]$. 

Now we follow the technique introduced in Ref.~\cite{curty2009} (see also Appendix~\ref{ApDecoy}). In particular, for each pair of values $n$ and $m$, we search for the parameter $\lambda_{nm}^c$ (which we shall call $\lambda^{nmc}_{\text{BSA}}$) and the entangled state $\rho^{nm}_{\text{ent},c}$ which correspond to the BSA of the states $\sigma_{AB,c}^{nm}\in{}S^{nm}_c$. That is, 
\begin{equation}
\lambda^{nmc}_{\text{BSA}}=\max\{\lambda_{nm}^c|\sigma_{AB,c}^{nm}\in S^{nm}_c\}. 
\end{equation}
Then we have that the secret key rate is upper bounded by
\begin{equation}\label{RateUpMDI}
R_U\leq\sum_{c\in S_{\text{an}}}\sum_{n,m\geq 1}p_{c|nm}r_{nm}(1-\lambda_{\text{BSA}}^{nmc})I_{nm,c}^{\text{ent}}(A;B),
\end{equation}
where $p_{c|nm}$ is the conditional probability that Charles announces $c$ given that Alice (Bob) sends him an $n$-photon ($m$-photon) state, $r_{nm}\approx e^{-2\mu}\mu^{n+m}/(n!m!)$ is the probability that Alice and Bob send Charles an $n$-photon state and an $m$-photon state, respectively, where $\mu$ is the mean photon number of their WCPs, and $I_{nm,c}^{\text{ent}}(A;B)$ is the Shannon mutual information calculated on the statistics $q_{kj}^{nmc}=\Tr(A_k\otimes B_j \rho_\text{ent,c}^{nm})$, with $\rho_\text{ent,c}^{nm}$ being the entanglement part of the BSA of the states $\sigma_{AB,c}^{nm}\in{}S^{nm}_c$.

To calculate $\lambda_{\text{BSA}}^{nmc}$ and the corresponding entangled state $\rho^{nm}_{\text{ent},c}$ for the BSA we use again SDP. For this, note that Alice's (Bob's) measurement operators $\{A_k\}_k$ ($\{B_j\}_j$) can be described by a projective measurement in a four-dimensional Hilbert space, {\it i.e.,}\ $A_k=\dyad{k}{k}$ ($B_j=\dyad{j}{j}$) with $k\in\{1,2,3,4\}$ ($j\in\{1,2,3,4\}$). Each operator $A_k$ ($B_j$) is associated with Alice (Bob) sending one of the four possible polarization states of the BB84 protocol to Charles.

In addition, and similar to the case of Appendix~\ref{ApDecoy}, we have that both the reduced density matrices of Alice and Bob are fixed by their state preparation processes. More precisely, $\rho_A^{nm}=\Tr_B(\sigma^{nm}_{AB})$ and $\rho_B^{nm}=\Tr_A(\sigma^{nm}_{AB})$ are both equal to Eq.~(\ref{ReducedMatrix}), where $\sigma^{nm}_{AB}=\sum_{c\in S_{\text{an}}} p_{c|nm}\sigma^{nm}_{AB,c}$. In fact, in this case, these conditions can even be generalized to $\sigma^{nm}_{AB}=\rho_A^{n}\otimes\rho_B^{m}$.

Putting all the conditions together, one can obtain the parameter $\lambda_{\text{BSA}}^{nmc}$ and the corresponding entangled state $\rho_{\text{ent,}c}^{nm}$, for each $n$, $m$ and $c$, by solving the following SDP,
\begin{eqnarray}
\min &\text{ }& 1-\Tr(\sigma^{nm}_{\text{sep},c}(\bm x)),\nonumber\\
\text{s.t.} && \sigma^{nm}_{AB,t}(\bm x)\geq 0 \text{ }\forall t\in S_{\text{an}},\nonumber\\
&&\Tr[\sigma^{nm}_{AB,t}(\bm x)]=1 \text{ }\forall t \in S_{\text{an}},\nonumber\\
&&\Tr[A_k\otimes B_j\sigma^{nm}_{AB,t}(\bm x)]=p_{kj}^{nmt}, \forall k,j,\forall t\in S_{\text{an}}\nonumber\\
&&\sum_{t\in S_{\text{an}}} p_{t|nm}\sigma^{nm}_{AB,t}(\bm x)=\rho_A^{n}\otimes\rho_B^{m},\nonumber\\
&& \sigma^{nm}_{\text{sep},c}(\bm x) \geq 0, \nonumber\\
&& \sigma^{nm,\Gamma}_{\text{sep},c}(\bm x) \geq 0, \nonumber\\
&& \sigma^{nm}_{AB,c}(\bm x)-\sigma^{nm}_{\text{sep},c}(\bm x) \geq 0,\label{MDI_Const}
\end{eqnarray}
where, as mentioned previously, we disregard for simplicity the secret key coming from positive partial transposed entangled states~\cite{horodecki2005} by neglecting in Eq.~(\ref{MDI_Const}) the key material provided  by those states $\sigma^{nm}_{\text{sep},c}(\bm x)$ that satisfy $\sigma^{nm,\Gamma}_{\text{sep},c}(\bm x) \geq 0$. A general but computationally more demanding method that considers also the key provided by positive partial transposed entangled states has been proposed, for instance, in Ref.~\cite{moroder2006}. Let $\bm x_{sol}$ denote the solution given by the SDP in~Eq.~(\ref{MDI_Const}), then
\begin{eqnarray}
\lambda^{nmc}_{\text{BSA}}&=&\Tr(\sigma^{nm}_{\text{sep},c}(\bm x_{sol})),\nonumber\\
\rho_{\text{ent},c}^{nm}&=&\frac{\sigma^{nm}_{AB,c}(\bm x_{sol})-\sigma^{nm}_{\text{sep},c}(\bm x_{sol})}{1-\lambda_{\text{BSA}}^{nmc}}.
\end{eqnarray} 

\end{document}